# Plasmonic antennas and zero mode waveguides to enhance single molecule fluorescence detection and fluorescence correlation spectroscopy towards physiological concentrations

Deep Punj, Petru Ghenuche, Satish Babu Moparthi, Juan de Torres, Victor Grigoriev, Hervé Rigneault, Jérôme Wenger

CNRS, Aix Marseille Université, Centrale Marseille, Institut Fresnel, UMR 7249, 13013 Marseille, France

**Abstract**

Single-molecule approaches to biology offer a powerful new vision to elucidate the mechanisms that underpin the functioning of living cells. However, conventional optical single molecule spectroscopy techniques such as Förster fluorescence resonance energy transfer (FRET) or fluorescence correlation spectroscopy (FCS) are limited by diffraction to the nanomolar concentration range, far below the physiological micromolar concentration range where most biological reaction occur. To breach the diffraction limit, zero mode waveguides and plasmonic antennas exploit the surface plasmon resonances to confine and enhance light down to the nanometre scale. The ability of plasmonics to achieve extreme light concentration unlocks an enormous potential to enhance fluorescence detection, FRET and FCS. Single molecule spectroscopy techniques greatly benefit from zero mode waveguides and plasmonic antennas to enter a new dimension of molecular concentration reaching physiological conditions. The application of nano-optics to biological problems with FRET and FCS is an emerging and exciting field, and is promising to reveal new insights on biological functions and dynamics.

**Introduction**

There is a strong hope that the next decades will see the emergence of personalized DNA sequencing and high-throughput screening for pathogens at affordable cost and viable time.[1,2] This quest is deeply related to the ability to interrogate and monitor individual molecules. Indeed, one of ultimate goals in life sciences and biotechnology is to observe how single molecules work and interact in their native physiological environment.[2,3] The single-molecule approach bears the intrinsic advantage to reveal information not normally accessible by ensemble measurements, such as sample heterogeneity, local concentration, and variances in kinetic rates. It does not require any perturbing synchronization of molecules to reach a sufficient ensemble-averaged signal, and it circumvents the need for 100% pure samples. Complex problems, such as protein structure folding, molecular motor operation or single-nucleotide polymorphism detection, are best studied at the single molecule level because of the molecular structure dispersion and the stochastic nature of the processes.[4] Although modern molecular biology has made enormous progress in identifying single molecules and their functions, efficiently detecting a single molecule is still a major goal with applications in chemical, biochemical and biophysical analysis. Progress towards this goal crucially depends on the development of techniques that provide visualisation and imaging of processes down to the molecular scale in intact cells.

The common strategy to optical single molecule fluorescence detection using Single Particle Tracking (SPT), Förster fluorescence resonance energy transfer (FRET) or fluorescence correlation spectroscopy (FCS) is bound to confocal microscopy [2] or total internal reflection fluorescence (TIRF) microscopy.[5-7] Despite their high sensitivities, these approaches are restricted to experimental conditions of low fluorophore density. To achieve single molecule analysis, the microscopic observation volume must only host a single fluorescent molecule of interest during the measurement acquisition time. Diffraction-limited optics generate detection volumes on the order of 0.5 fL, which imposes concentrations of the fluorescent species in the nanomolar range so as to isolate a single molecule in the detection volume. However, most biologically relevant processes, such as transient interactions between proteins and nucleic acids or between enzymes and their ligands, calls for ligand concentrations in the micro to millimolar range to ensure reaching relevant reaction kinetics and biochemical stability (see Figure 1). Unfortunately, this size regime is not accessible by classical diffraction-limited optical microscopy techniques. To be able to investigate processes down to the single molecule level, the detection volume must be reduced by at least three orders of magnitude as compared to confocal microscopy.[8-14] Besides this challenge, the diffraction phenomenon ultimately limits the amount of collected light from a single molecule and the achievable signal-to-background ratio which actually determines the maximum acquisition speed and temporal resolution of the experiments. As a consequence, single molecule detection can be performed only on fluorescent species with high quantum yields and good photostability.

To breach the diffraction limitations, plasmonic antennas are promising tools to control and manipulate optical fields down to the nanometre scale.[9] Plasmonic antennas are devices that convert freely propagating optical radiation into localized energy, and vice versa, in close analogy to their classical radiowave antenna counterparts.[15-17] Plasmonic antennas exploit the unique optical properties of metallic nanostructures that support collective electron excitations, known as surface plasmons. Their ability to achieve extreme light concentration unlocks an enormous potential of application to enhance single molecule fluorescence detection. Although resistive heating losses in metals can severely limit the performance of plasmonics antennas and lead to dramatic fluorescence quenching, many useful functionalities have recently been realized despite the presence of loss.[15-17] Moreover, plasmonic quenching losses critically depend on the distance from the fluorescent molecule to the antenna and its spectral emission properties respective to the antenna resonances. Therefore, quenching losses to the metal can be controlled to some extent by carefully designing the antenna, chemically functionalising the metal surface, and/or selecting an appropriate dye emitter.[18-20]

In this review, we discuss current physics-based strategies employed to break the concentration barrier and enhance the fluorescence detection of single molecules. Recently, a review paper has been published on the different physical and biochemical approaches to overcome the concentration barrier for the detection of fluorescent molecules towards low and high concentrations.[13] Here, we discuss in more details the physical approaches and their limitations, as the physical approaches are the most versatile and promising techniques to improve the single molecule fluorescence detection towards high concentrations. We focus first on metal nanoapertures or zero-mode-waveguides (ZMW) that pushed the scientific field ahead over the last decade. Then we review recent progress taking advantages of plasmonic antennas and near-field scanning optical microscope (NSOM) probes. The differences behind the approaches to plasmonic antennas are better revealed by considering the nanofabrication strategy employed: either by using high-end nanofabrication tools such as electron beam lithography of focused ion beam (top-down approach), or by using chemical synthesis and self-assembly (bottom-up approach).

## ZERO-MODE WAVEGUIDES AND PLASMONIC NANOAPERTURES

**Nanoscale concentration of light**

Using single nanometric apertures milled in an opaque metallic film is a conceptually simple, flexible and robust method to generate a detection volume much below the diffraction barrier and enable single molecule analysis at higher concentrations. This concept was introduced in 2003 by the groups of Harold Craighead and Watt Webb in a pioneering contribution.[21] The nanoaperture acts as a pinhole directly inserted into the microscope sample plane (Fig. 2a,b). As the aperture diameter is reduced below half of the optical wavelength, the light inside the aperture is confined to a rapidly decaying evanescent mode, with a decay length of a few tens of nanometers (Fig. 2c). Subwavelength apertures have thus been termed zero-mode waveguides (ZMW) to emphasize the evanescent nature of the excitation light inside the aperture. With a typical diameter of 100nm, a single nanoaperture enables reaching a detection volume of about 2 attoliters ($10^{-18}$L), which is over three orders of magnitude smaller than diffraction-limited confocal volumes (Fig. 2e).[22,23]

A second major effect brought by the sub-wavelength aperture is that it can significantly enhance the detected fluorescence rate per emitter, thereby increasing the signal-to-noise ratio for single molecule detection. Using single rhodamine 6G molecules in isolated 150nm diameter apertures milled in aluminum, a 6.5 fold enhancement of the fluorescence rate per molecule was reported as compared to free solution.[22] Further enhancement up to 25-fold can be obtained by tuning the plasmon properties of the nanoapertures.[24-28] The physical phenomena leading to the fluorescence enhancement in single subwavelength apertures were clarified in reference [29] as a result of gains in excitation intensity, quantum yield and fluorescence collection efficiency. The evolution of the fluorescence enhancement with the aperture diameter (Fig. 2f) also shows that fluorescence quenching becomes predominant for aperture diameters below 100nm, as a consequence of nonradiative energy transfer to the free electrons of the metal. This explains the existence of an optimum diameter to maximize fluorescence enhancement. The high fluorescence brightness per emitter achieved with the nanoapertures not only improves the signal quality, it also enables performing faster integration times (Fig. 2d).[30] This opens the way to monitoring biochemical reactions at reduced time scales, which fits well the time resolution requirements at high ligand concentrations.

**Enhanced single molecule detection in solution with zero mode waveguides**

A large range of biological processes have been monitored with single molecule resolution at micromolar concentrations while using ZMW nanoapertures. Most studies take advantage of fluorescence correlation spectroscopy (FCS) as a biophotonic method to analyse the fluorescence intensity trace from individual molecules diffusing inside and outside the nanoaperture (Box 1). As an example of the effectiveness of nanoapertures for performing single-molecule experiments at high concentrations, DNA polymerase activity has been monitored at 10 µM dye concentration with an average of 0.1 molecule inside a 43-nm-diameter aperture.[21] However, for experiments conducted on ultrasmall structures, the signal to noise ratio comes close to one, as a consequence of quenching losses and increased background.

The pioneering work of Levene and co-workers [21] has led to a number of studies combining nanometric apertures with single molecule detection. The applications include oligomerization of the bacteriophage λ-repressor protein,[35] protein-protein interactions considering the GroEL-GroES complex,[36-37] or observation of flow mixing.[38] The applications can be extended to dual-color cross-correlation FCCS analysis to monitor DNA enzymatic cleavage at micromolar concentrations with improved accuracy.[39] To avoid the use of fluorescent labelling, the fluorescence detection technique can be operated in reverse mode: the solvent solution filling the aperture is made highly fluorescent by using a millimolar concentration of small fluorescent molecules. Label-free (non-fluorescent) analytes diffusing into the aperture displace the fluorescent molecules in the solution, leading to a decrease of the detected fluorescence intensity, while analytes diffusing out of the aperture return the fluorescence level.[40]

A very promising application of nanometric apertures concerns real-time single-molecule DNA and RNA sequencing.[41-43] The development of personalized quantitative genomics requires novel methods of DNA sequencing that meet the key requirements of high-throughput, high-accuracy and low operating costs simultaneously. To meet this goal, each nanoaperture forms a nano-observation chamber for watching the activity of a single DNA polymerase enzyme performing DNA sequencing by synthesis (Fig. 3).[41] The sequencing method records the temporal order of the enzymatic incorporation of the fluorescent nucleotides into a growing DNA strand replicate. Each nucleotide replication event lasts a few milliseconds, and can be observed in real-time. Currently, over 3000 nanoapertures can be operated simultaneously, allowing massive parallelization.

The applicability of ZMW is deeply related to the ability to bind molecules of interest in the subdiffraction observation volume without interfering with their biological function. However, nonspecific adsorption to the metal cladding remains a critical limitation. To address this issue, metal passivation protocols have been specifically designed for aluminum using polyphosphonate chemistry,[44] and gold using methoxy-terminated, thiol-derivatized polyethylene glycol.[45] These functionalization schemes enable target biomolecules to be selectively tethered to the silica bottoms of nanoapertures.

**Investigating live cell membranes at the nanometer scale with zero mode waveguides**

Many cell membrane receptors involved in cell communication have dimensions on the order of tens of nanometres, much below the diffraction-limited optical resolution of classical microscopes. Probing the cell membrane organization with nanometer resolution is a challenging task, as standard optical microscopy does not provide enough spatial resolution while electron microscopy lacks temporal dynamics and cannot be easily applied to live cells. Nanoapertures combined to FCS offer the advantages of both high spatial and temporal resolution together with a direct statistical analysis (Fig. 4a). The nanoaperture works as a pinhole directly located under the cell to restrict the illumination area (Fig. 4b). Diffusion of fluorescent markers incorporated into the cell membrane provide the dynamic signal, which is analyzed by correlation spectroscopy to extract information about the membrane organization (Fig. 4c,d).[46-48] To gain more insight about the membrane organization, measurements can be performed with increasing aperture diameters.[49] For instance, it was shown that fluorescent chimeric ganglioside proteins partition into 30nm structures inside the cell membrane. Apart from the translational diffusion, the stoichiometry of nicotinic acetylcholine and P2X2 ATP receptors isolated in membrane portions inside zero-mode waveguides was analyzed using single-step photobleaching of green fluorescent protein incorporated into individual subunits.[50]

A key requirement for these methods is the need for cell membranes to adhere to the substrate. Cell membrane invagination within the aperture was shown to depend on the membrane lipidic composition,[47] and on actin filaments.[51] To further ease cell adhesion, and avoid membrane invagination issues, planarized 50nm diameter apertures have been recently introduced.[52] The planarization procedure fills the aperture with fused silica, to achieve no height distinction between the aperture and the surrounding metal.

**Extra plasmonic control of the fluorescence directivity**

Due to its subwavelength dimension, an isolated nanoaperture does not provide a strong directional control on the light emitted from the aperture. Adding concentric surface corrugations (or grooves) opens new possibilities to control the fluorescence directionality (Fig. 5), while preserving the light localization inside the nanoaperture. The corrugations have two main roles. First, when the corrugations are milled on the surface receiving the excitation beam (reception mode), the grating formed by the corrugations provides the supplementary momentum required to match the incoming light to surface-plasmon polariton modes, which further increase the light intensity at the central aperture. Second, when the corrugations are milled on the surface where the fluorescence is collected (emission mode), the reverse phenomenon appears, the surface corrugations couple the surface waves back to radiated light into the far-field. As the coupling of far-field radiation into SPP modes is governed by geometrical momentum selection rules, the coupling occurs preferentially at certain angles for certain wavelengths.

Corrugated aperture have been reported to provide high fluorescence enhancement together with beaming of the fluorescence light into a narrow cone.[53,54] The fluorescence light from single molecules can thus be efficiently collected with a low numerical aperture objective, releasing the need for complex high NA objectives. By tuning the geometrical properties of the corrugation design, the fluorescence directionality can be controlled,[55,56] which offers photon sorting abilities from nanoscale volumes. Lastly, to release the need for complex nanofabrication, a new strategy has been presented, where the shallow grooves are replaced by nanoapertures milled into a regular array.[57]

> **Box 1: Fluorescence correlation spectroscopy**
>
> Fluorescence correlation spectroscopy (FCS) is a powerful and versatile method to analyse the fluorescence time trace from a single molecule diffusing in solution.[31-33] FCS is based on the statistical analysis of the temporal fluctuations affecting the fluorescence intensity by computing the second order correlation of the fluorescence intensity time trace. FCS can in principle provide information about any molecular process that induces a change in the fluorescence intensity. For instance, fluctuations occur when molecules diffuse in and out of an observation volume, or when reaction kinetics or conformational changes induce a change in the fluorescence brightness. Applications include determining translational and rotational diffusion, hydrodynamic radii, molecular concentrations, chemical kinetics, and binding reaction rates.[34]

**PLASMONIC OPTICAL ANTENNAS**

Sharp metal tips and edges can hold high local electromagnetic intensities due to the lightning rod effect. Hence a large part of plasmonic research has been devoted to the optimization and application of this local intensity enhancement phenomenon.[15-17] In terms of antenna concepts, a single circular nanoaperture such as a ZMW is simple to implement and robust, yet it appears far from being optimum. Another simple design to reduce the detection volume in single molecule fluorescence studies exploits the subwavelength confinement of light supported by surface plasmons at a metal-water interface.[58-60] In this method that appears as the plasmonic avatar of TIRF, the sample is a thin metal film deposited on a glass substrate that is illuminated at the surface plasmon resonance angle. The height of the detection volume is typically around 50nm, creating detection volumes on the order of 5-10aL. More advanced plasmonic optical antenna designs have been developed over the last years to achieve higher intensity enhancement and light confinement. In this section, we review these approaches and their applications to detect fluorescent biological molecules in solution.

**Near-field scanning optical microscope probes**

An intrinsic limitation of nanoapertures and planar antennas is that they are fixed on a substrate and hence unable to provide an image of the sample with nanometer resolution. True nanoscale microscopy can be achieved by attaching the aperture or antenna to a tapered optical fiber or cantilever probe and raster-scanning the device across a surface at a distance of a few nanometers above the sample to provide an image. This technique has been termed NSOM for near-field scanning optical microscopy.[61,62]

A standard approach to NSOM probes implements tapered optical single-mode fibers that are coated with metal. At the apex of the tip, an aperture of nanometer dimension is opened by focused ion beam milling. The nanoaperture at the apex of the tip constrains the illumination along both lateral and longitudinal directions, in similar fashion to zero mode waveguides. The light confinement can be used to improve the optical resolution for bioimaging, reaching about 50nm for imaging on cell membranes.[63-65] Dynamic FCS measurements with aperture-based NSOM probes have been reported on lipid bilayers,[66] single nuclear pore,[67] and intact living cell membranes.[68] These dynamic measurements provide sub-millisecond temporal resolution at spatial resolutions below 100nm. Using more advanced nanoaperture design such as a bowtie aperture[69] should improve the light throughput, enabling even better spatial and temporal resolution. Another approach uses gold nanoparticles attached to glass tips as NSOM probes.[18,19] Single calcium channels on erythrocyte plasma membranes have been visualized using a nanoparticle-based NSOM probe with a 50nm spatial resolution.[70]

The aperture-based and nanoparticle-based NSOM approaches can be combined into an elegant system where a resonant optical antenna tip is carved on top of a nanoaperture NSOM probe (Fig. 6).[9,71,72] This technique combines the background suppression from the aperture-based NSOM with the high local fields enhancement of the antenna tip. Moreover, the antenna tip can be used to control the polarization and direction of the molecular fluorescence emission.[71] Spatial resolutions below 30nm were achieved on single molecules fixed on a substrate in aqueous conditions as well on cell membranes in physiological conditions with virtually no background.[72] The method allows individual proteins to be distinguished from nanodomains, and the degree of clustering can be quantified by measuring the physical size and brightness of fluorescent spots.

**Top-down approaches to plasmonic antennas**

Top-down nanofabrication techniques can be envisioned as sculpting to remove the excess material where it is not needed. Typical techniques involve electron beam lithography, focused ion beam milling, or deep UV photolithography. These techniques have the potential to create almost any planar antenna design, and set the antennas to well defined locations, so that parallel measurements can be conducted by monitoring several antennas.

A remarkable example of plasmonic antenna to enhance single molecule fluorescence is shown in Fig. 7a.[20] The antenna consists of two facing gold nanotriangles fabricated by electron beam lithography, and is commonly referred to as bowtie antenna. Thanks to coupling between the localized surface plasmons from the two closely opposed nanotriangles, the excitation light is strongly confined inside the 10nm gap between the triangles with a local excitation intensity enhancement around 100. Despite non-radiative ohmic losses to the metal, the quantum yield of a near-infrared fluorescent dye is increased by ten times, from 2.5% to about 25%. The gain in local excitation intensity and fluorescence quantum yield combine to reach overall fluorescence enhancement factors per single molecule of three orders of magnitude that come simultaneously with fluorescence lifetime reductions down to 10ps. Experiments to extend FCS towards micromolar concentrations with bowtie antennas have been reported in [73]. The bowtie antenna resonance in the near infrared imposes to use fluorescent dyes that emit into the near IR. For the two fluorophores probed,[74] the bowtie-FCS signal was found dominated by molecules that transiently stick to the substrate near the bowtie gap, and by photobleaching/photoblinking dynamics on tens of millisecond time scale, much larger than typical translational diffusion times.

An elegant method to fabricate gold bowtie arrays with well-defined nanometer gaps has been reported by a combination of colloid lithography and plasma processing (Fig. 7b).[74] Controlled spacing of the antenna gap is achieved by taking advantage of the melting between polystyrene particles at their contact point during plasma processing and using this polymer thread as a mask for gold deposition. A supported lipid membrane can be formed on the intervening substrate by vesicle fusion, and diffusion trajectories of individual proteins are traced as they sequentially pass through multiple gaps where fluorescence enhancement takes place.

Plasmonic antennas appear as efficient tools to provide large enhancement of the fluorescence excitation and emission rates,[18-20] and direct the fluorescence light.[71,75] However, applications of plasmonic antennas to detect fluorescent molecules in solution of micromolar concentration are challenged by the large contribution in the detected fluorescence intensity from non-enhanced molecules tens of nanometer away from the antenna. In any nanoantenna experiment on molecules in solution, the observed fluorescence signal is a sum of two contributions: the enhanced fluorescence from the few molecules in the antenna gap region (hot spot) superimposed on a potentially large fluorescence background from the several thousands of molecules that are still present within the diffraction-limited confocal volume. To address this challenge, a design termed "antenna-in-box" has been proposed (Fig. 7c).[76] It combines a gap-antenna inside a nano-aperture, as is especially designed for enhanced single molecule analysis in solutions at high concentrations. The different components of the antenna-in-box have complementary roles: the gap-antenna creates the hot spot for fluorescence enhancement, while the surrounding nanoaperture screens out the background fluorescence from the molecules diffusing away from the central gap antenna. This design led to dramatic fluorescence enhancement factors above 1000 fold together with detection volumes down to 60 zeptoliters (Fig. 7d,e), enabling single molecule operation at concentrations above 20μM.

**Bottom-up approaches to plasmonic antennas**

Complementary to top-down nanolithography, bottom-up approaches to nanofabrication are based on the directed self assembly of atoms, molecules and/or nanoparticles into the desired nanostructure. Thanks to their low intrinsic cost, bottom-up approaches are promising for large scale applications of plasmonic nanostructures. As a prominent example of bottom-up nanofabrication, the synthesis of complex shaped metal nanoparticles has received a large attention.[77,78] Metal nanoparticles are largely available at a low intrinsic cost. They support local surface plasmon resonances that confine and enhance the electromagnetic fields within a few tens of nanometres close to the nanoparticle surface. Moreover, the nanoparticle spectral response can be tuned by selecting appropriate material and shape. In the context of fluorescence applications to detect biomolecules at physiological concentrations, metal nanoparticles appear naturally as an attractive nanodevice to overcome the diffraction limit for the concentration of light (Fig. 8a).[79-83] However, as for top-down lithographied plasmonic antennas, the detection of diffusing molecules at micromolar concentrations is challenged by the large contribution in the fluorescence signal from unenhanced molecules tens of nanometer away from the nanoparticles.[80,81] Molecular sticking to the metal may also become an issue to analyse the signal dynamics.[79,82] These issues can be avoided by using emitters with low quantum yield to take advantage of the higher fluorescence enhancement factors obtained with them,[82] or by using a chemical quencher to the solution so as to reduce the fluorophore's quantum yield and maximize the fluorescence enhancement.[83] The use of surfactant and salts in the solution was also found to reduce the binding of molecules to the gold surface.[76,83] With these precautions, the near-field detection volume and average fluorescence enhancement set by a single gold nanoparticle were quantified,[83] with detection volumes down to 270 zeptoliters (three orders of magnitude beyond the diffraction barrier) together with 60 fold enhancement of the fluorescence brightness per molecule (Fig. 8b). It should be noted that significantly higher enhancement factors can be reached by selecting nanoparticles with sharper plasmonic resonances such as nanorods, for which enhancement factors up to 1000 fold have been reported.[82] Additionally, silver nanoisland films prepared by wet chemical synthesis or thermal vapor deposition benefit from simple nanofabrication technique and have been reported to enhance FCS application up to 9 μM concentration.[84]

To provide tighter confinement of light and larger fluorescence enhancement factors, nanoantennas can benefit from the electromagnetic coupling between several nanoparticles separated by distances much smaller than the nanoparticles radii (Fig. 8e). Gold nanoparticle dimers linked by a single DNA double-strand can be synthesized and filtered using electrophoresis (Fig. 8c).[85-87] The interparticle distance defining the antenna gap is tuned by changing the length of the DNA template, and a binding site to target a single fluorescent molecule can be inserted in the structure. To provide further flexibility in the design of plasmonic antennas, DNA origami is a powerful method to obtain excellent nanofabrication control.[88-90] Gold nanoparticles with diameters up to 100nm were attached to DNA origami pillar structures, reaching gaps of 23nm which also incorporated docking sites for fluorescent molecules (Fig. 8d).[90] Thanks to the large scattering cross-section of these antennas and the operation near resonance, a maximum of 117 fold fluorescence enhancement was obtained for a single ATTO647N fluorescent molecule (Fig. 8f). Thanks to the intensity enhancement introduced by the nanoantenna, single-molecule measurements could be performed at concentrations up to 500nM, two orders of magnitude higher than conventional measurements.[90,91] Plasmonic antennas templated with DNA origami open the way for the development of bottom-up inexpensive enhancement chambers for biological assays with single molecule resolution at high physiological concentrations.

**Conclusion**

Monitoring single molecules at the physiologically relevant micromolar concentration regime imposes to rethink the optical microscope apparatus to break the diffraction limit. This difficulty can be accounted as one of the main limitations for the broad applicability of optical single-molecule detection in biology and medicine.[9,13] The ability to reliably fabricate nanostructures to confine and enhance the light into nanoscale volumes paves the way to overcome the diffraction challenge, and several methods based on zero-mode waveguides or plasmonic antennas have been reviewed here. Moreover, the plasmonic approach can benefit from other approaches using advanced microscopy techniques,[92,93] dielectric-based antennas,[94-98] microfluidics,[99,100] or optical fibre probes.[101-103] All these techniques, and their combination, significantly expand the single molecule toolbox. The application to biological problems is an emerging and exciting field, which is promising to reveal new insights on biological functions and dynamics.

**Acknowledgments**

The authors acknowledge stimulating discussions with M. F. Garcia-Parajo, N. F. Van Hulst and T. W. Ebbesen. This work has received funding from the European Research Council under the European Union's Seventh Framework Programme (FP7/2007-2013) / ERC Grant agreement 278242 (ExtendFRET) and 288263 (NanoVista).

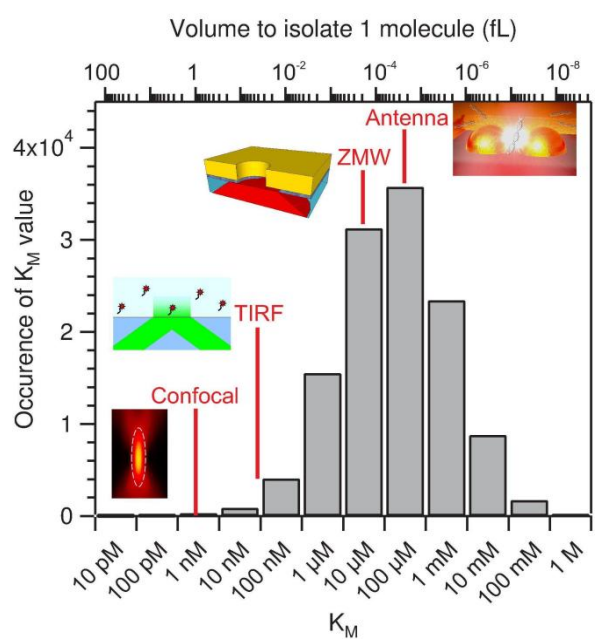

**Figure 1.** Ultrasmall detection volumes are needed to investigate enzymatic function at the single molecule level. Histogram of Michaelis constant $K_M$ for 118,000 enzymes taken from the Brenda database (http://www.brenda-enzymes.org/) in November 2013. The top axis shows the detection volume required to isolate a single molecule. The vertical bars indicate the effective concentration regime and detection volume reached by different techniques (TIRF: total internal reflection fluorescence microscopy; ZMW: zero mode waveguide).

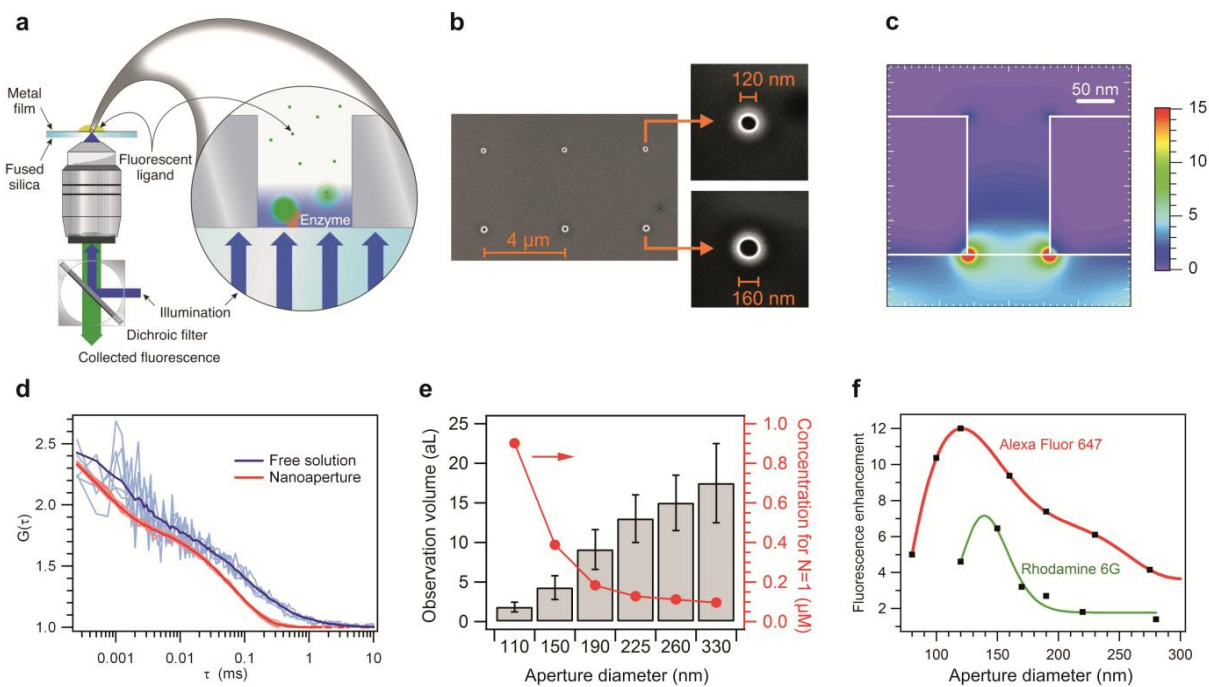

**Figure 2.** Zero-mode waveguides and nanoapertures to enhance the detection of single fluorescent molecules at micromolar concentrations. (a) Nanoaperture for enhanced single molecule fluorescence detection at micromolar concentrations.[21] (b) Electron microscope images of 120 and 160 nm apertures milled in gold.[28] (c) Field intensity distribution on a 120 nm water-filled gold aperture illuminated at 633 nm.[28] The spatially-averaged excitation intensity over the nanoaperture detection volume as seen by FCS is enhanced about three times as compared to the diffraction-limited confocal spot. (d) Comparison of normalized FCS correlation traces between confocal and nanoaperture configurations: to reach similar amplitudes, the concentration was increased by a factor 400 for the nanoaperture. Moreover, the nanoaperture enables observing short diffusion times with significantly improved signal to noise ratio.[30] (e) Observation volumes measured for aluminum apertures. The right axis shows the corresponding concentration to ensure there is a single molecule in the observation volume.[22] (f) Fluorescence brightness enhancement factor for Alexa Fluor 647 molecules in apertures milled in gold (laser excitation 633nm) [25] and for Rhodamine 6G molecules in apertures milled in aluminum (laser excitation 488nm). Figures reproduced with permission: (a) © AAAS 2003, (b,c) © ACS 2010, (d) © ACS 2009, (e) © APS 2005, (f) © APS 2008.

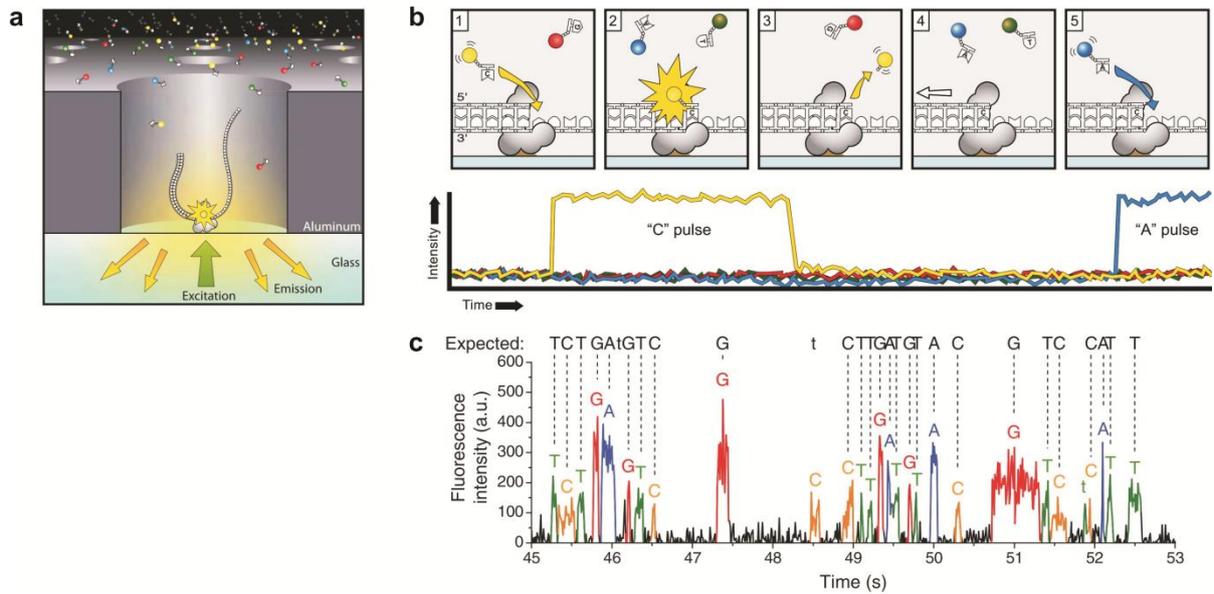

**Figure 3.** Application of zero-mode waveguides to single-molecule real-time DNA sequencing. (a) Principle of the experiment: a single DNA polymerase is immobilized at the bottom of a ZMW, which enables detection of individual phospholinked nucleotide substrates against the bulk solution background as they are incorporated into the DNA strand by the polymerase. (b) Schematic event sequence of the phospholinked dNTP incorporation cycle, the lower trace displays the temporal evolution of the fluorescence intensity. (c) Section of a fluorescence time trace showing 28 incorporations events with four color detection. Pulses correspond to the least-squares fitting decisions of the algorithm. Figures reproduced with permission [41] © AAAS 2009.

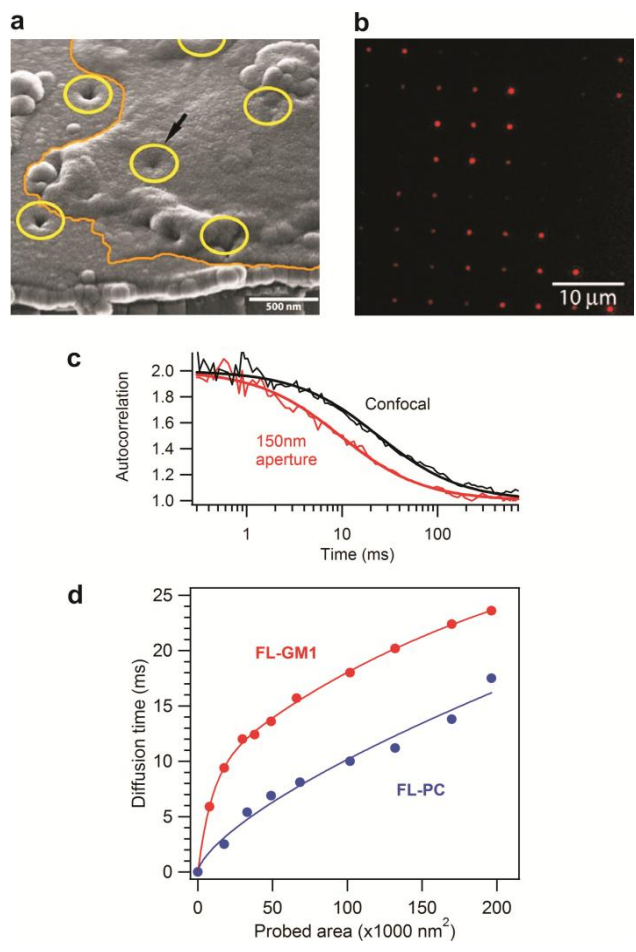

**Figure 4.** Application of zero-mode waveguides to investigate cell membranes below the diffraction limit. (a) Scanning electron microscope image of cross-sectional cuts of nanoapertures. Cell membranes have been outlined (lightgray), and aperture locations have been circled. Cell membrane spanning a nanoaperture dips down (arrow), suggesting membrane invagination. The scale bar is 500nm. (b) Fluorescence micrographs of cells labelled with DiI-C 12 membrane probe through 280nm aluminum apertures. (c) Normalized FCS correlation functions and numerical fits (thick lines) obtained for the FL-$G_{M1}$ ganglioside lipid analog, demonstrating a significant diffusion time reduction in the nanoaperture. (d) Molecular diffusion times versus aperture area for the FL-$G_{M1}$ ganglioside and FL-PC phosphatidylcholine. Figures reproduced with permission: (a,b) [51] © IOP 2007, (c,d) [49] © BS 2007.

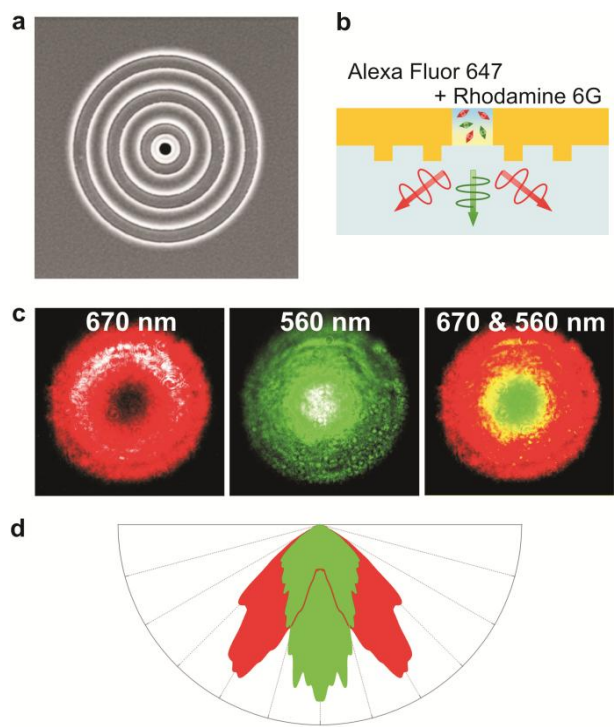

**Figure 5.** Corrugated nanoapertures to control the fluorescence directivity with surface plasmon waves. (a) Scanning electron microscope image of a single aperture of 140nm diameter milled in gold with two concentric grooves of period 440nm and depth 65nm. (b) Sketch of the experiment to illustrate the photon sorting ability: the central aperture is filled with a mixed solution of Alexa Fluor 647 and Rhodamine 6G. (c) Radiation patterns in the back focal plane of the objective for emission centered at 670 nm and 560 nm. (d) Fluorescence radiation pattern for the two different emission wavelengths. Figures reproduced with permission [55] © ACS 2011.

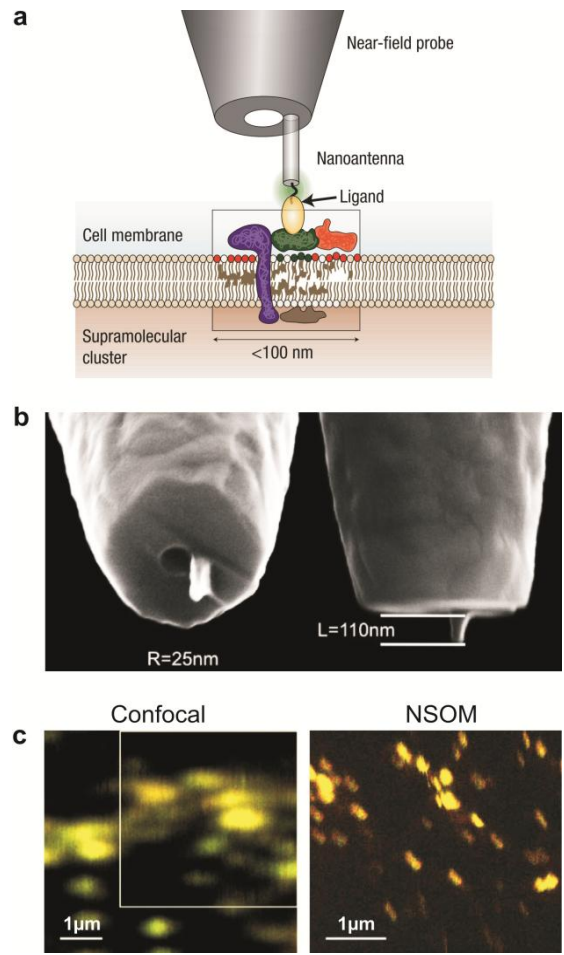

**Figure 6.** (a) Optical antenna carved on top of a NSOM aperture probe. Topography, biochemical recognition and fluorescence images can be recorded simultaneously at nanometre resolution.[9] (b) Scanning electron microscope image of a tip-on-aperture probe. (c) Zoomed-in confocal microscopy image of LFA-1 at the cell surface of monocytes visualized by confocal microscopy (left). The right panel shows the NSOM imaging of the highlighted region in the confocal image.[72] Figures reproduced with permission: (a) © NPG 2008, (b,c) © Wiley-VCH 2010.

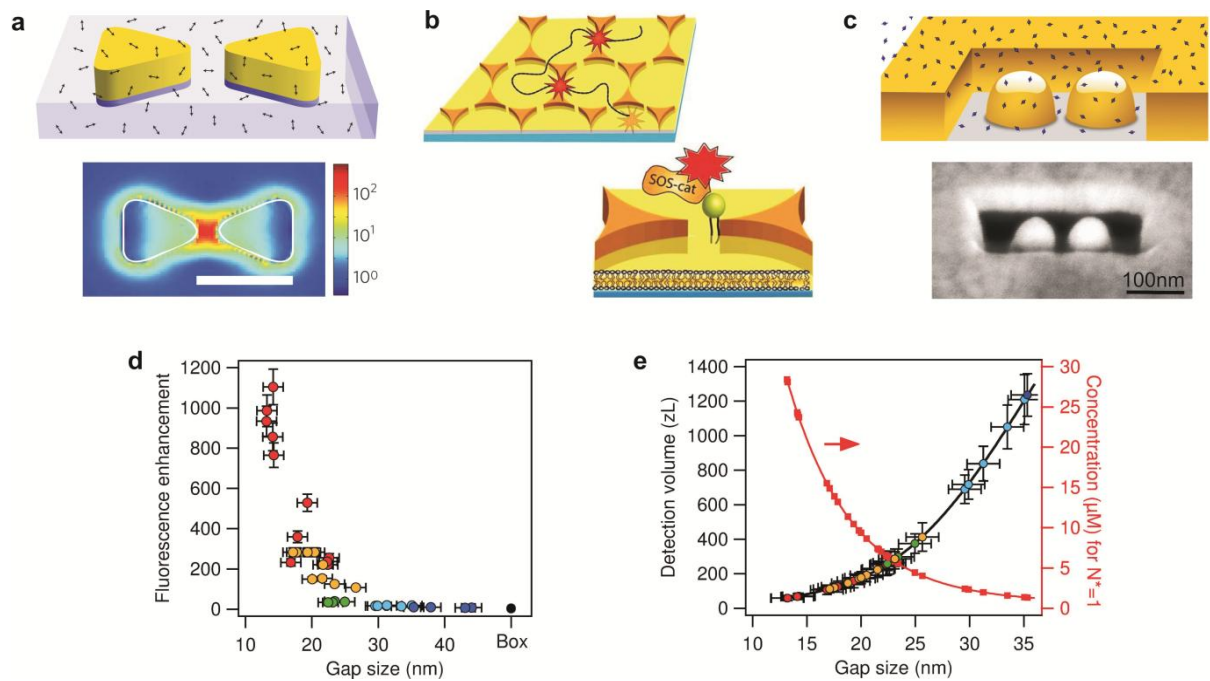

**Figure 7.** Top-down approaches to plasmonic antennas for enhanced single molecule fluorescence. (a) Gold bowtie antenna covered by fluorescent molecules (arrows) in PMMA resin. The lower image shows the computed local intensity enhancement. The scale bar is 100 nm.[20] (b) Plasmonic bowtie antennas surrounded by a fluid supported lipid bilayer, where fluorescently labeled Ras proteins are anchored in the upper leaflet of the lipid membrane. Fluorophores tethered to the supported membrane can diffuse in the plane and thereby pass through the nanogaps.[74] (c) Antenna-in-box platform for single-molecule analysis at high concentrations.[76] Scatter plots of the fluorescence enhancement factor (d) and measured detection volume (e) as a function of gap size for the antenna-in-box.[76] Figures reproduced with permission: (a) © NPG 2009, (b) © ACS 2012, (c-e) © NPG 2013.

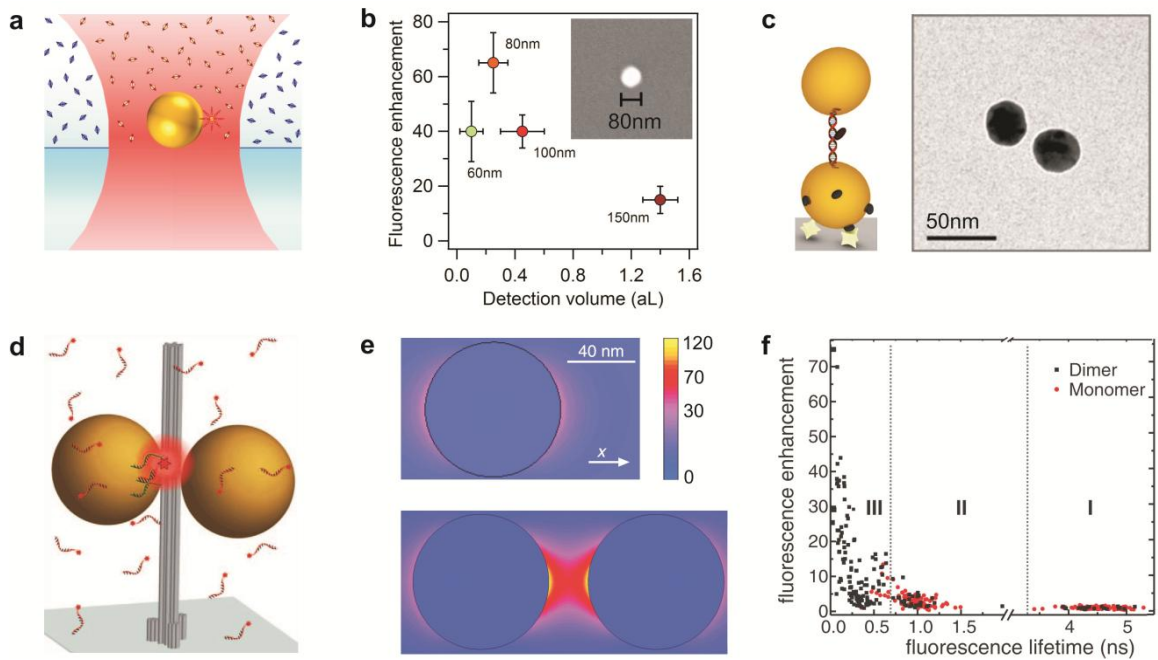

**Figure 8.** Bottom-up approaches to plasmonic antennas for enhanced single molecule fluorescence. (a) A single gold nanoparticle is used as optical antenna.[83] (b) Fluorescence enhancement versus the near-field detection volume obtained with single gold nanoparticles,[83] the nanoparticle diameter is annotated close to the data point. (c) Cryo-EM of a plasmonic dimer antenna made of two 40nm gold particles linked with a 30 base pairs double stranded DNA.[85] (d) DNA origami pillar with two gold nanoparticles forming a dimer antenna.[90] Fluorescent labeled ssDNA sequences in solution can transiently hybridize with complimentary sequences in the origami structure at the hotspot between the particles. (e) Numerical simulation of electric field intensity for single and dimer of 80nm diameter gold particles. The incoming light is horizontally polarized at a wavelength of 640 nm, the gap distance in the dimer is 23nm.[90] (f) Scatter plot of fluorescence intensity versus lifetime of the ATTO647N-labeled DNA origami pillar with binding sites for one (monomer) and two (dimer) 80-nm diameters particles. Figures reproduced with permission: (a) © OSA 2013, (c) © NPG 2012, (d-f) © AAAS 2012.